\documentclass[]{spie}

\usepackage{amsmath,amsfonts,amssymb}
\usepackage{graphicx}
\usepackage{scalerel}
\usepackage[colorlinks=true, allcolors=blue]{hyperref}

\title{Ion optical clocks with three electronic states}

\author{C. A. Holliman\thanks{\ \ cholliman3@gmail.com}}
\author{M. Fan}
\author{A. M. Jayich}
\affil{Department of Physics, University of California, Santa Barbara, California 93106, USA}

\begin{document} 
\maketitle

\begin{abstract}
Optical clocks are the apotheosis of precision measurement, but they require frequent maintenance by scientists.  The supporting laser systems are a particularly demanding component of these instruments.  To reduce complexity and increase robustness we propose an optical clock with trapped alkali-like ions that use the $S_{1/2}\rightarrow D_{3/2}$ electric quadrupole transition. Compared to traditional group-II ion clocks this reduces the number of laser wavelengths required, and uses hyperfine state preparation and readout techniques enabled by the nuclear spin $I=1/2$.  We consider $^{225}$Ra$^{+}$ as a candidate system for a clock with three electronic states, and discuss the potential to help realize a transportable optical clock.
\end{abstract}


\section{INTRODUCTION}
\label{sec:intro}

Transportable optical clocks are promising for tests of Einstein's equivalence principle \cite{Takamoto2020}, searches for dark matter \cite{Derevianko2014}, as well as improved timekeeping and global positioning systems \cite{Ludlow2015}.  In combination with terrestrial clock networks, optical clocks on satellites would improve limits on the accuracy of intercontinental frequency transfer and comparison, and aid deep space navigation \cite{Burt2021, BACON2021Nature}.  However, there is a gap between existing state-of-the-art optical clocks and the desired turn-key systems that can run autonomously, e.g. as part of an advanced global positioning system network.  Generation, control, and delivery of laser light with integrated photonics is desirable for such advanced clocks.  The use of integrated photonics is promising for achieving smaller clock form factors and system robustness, where optical alignment can be lithographically defined.  We propose trapped ion optical clocks that can operate with three or even two lasers at low powers and at wavelengths longer than 400 nm.  These features should reduce the barriers for realizing a clock with integrated photonics.

Clocks based on trapped ions that can be directly laser cooled are a potential path towards robust transportable optical clocks.  An ion's charge enables continuous trapping for months in an rf Paul trap, and clock operation only requires low-power lasers.  In realizations with candidate systems Ca$^{+}$ and Sr$^{+}$ \cite{Huang2020, Loh2020}, the excited $D_{5/2}$ clock state is separate from the states used for laser cooling and state detection, and an additional laser is needed to reset the electronic population between clock interrogations \cite{Madej2004, Chwalla2009}.  In total, five electronic states are used along with four laser wavelengths for these relatively simple clocks. To further ease requirements for the use of integrated photonics, the number of required laser wavelengths can be reduced from four to three, or possibly only two with optical clocks based on the $S_{1/2}\rightarrow D_{3/2}$ electric quadrupole (E2) transition of singly-charged alkali-like ions.

\begin{figure}
\centering{}
\includegraphics[height=6cm]{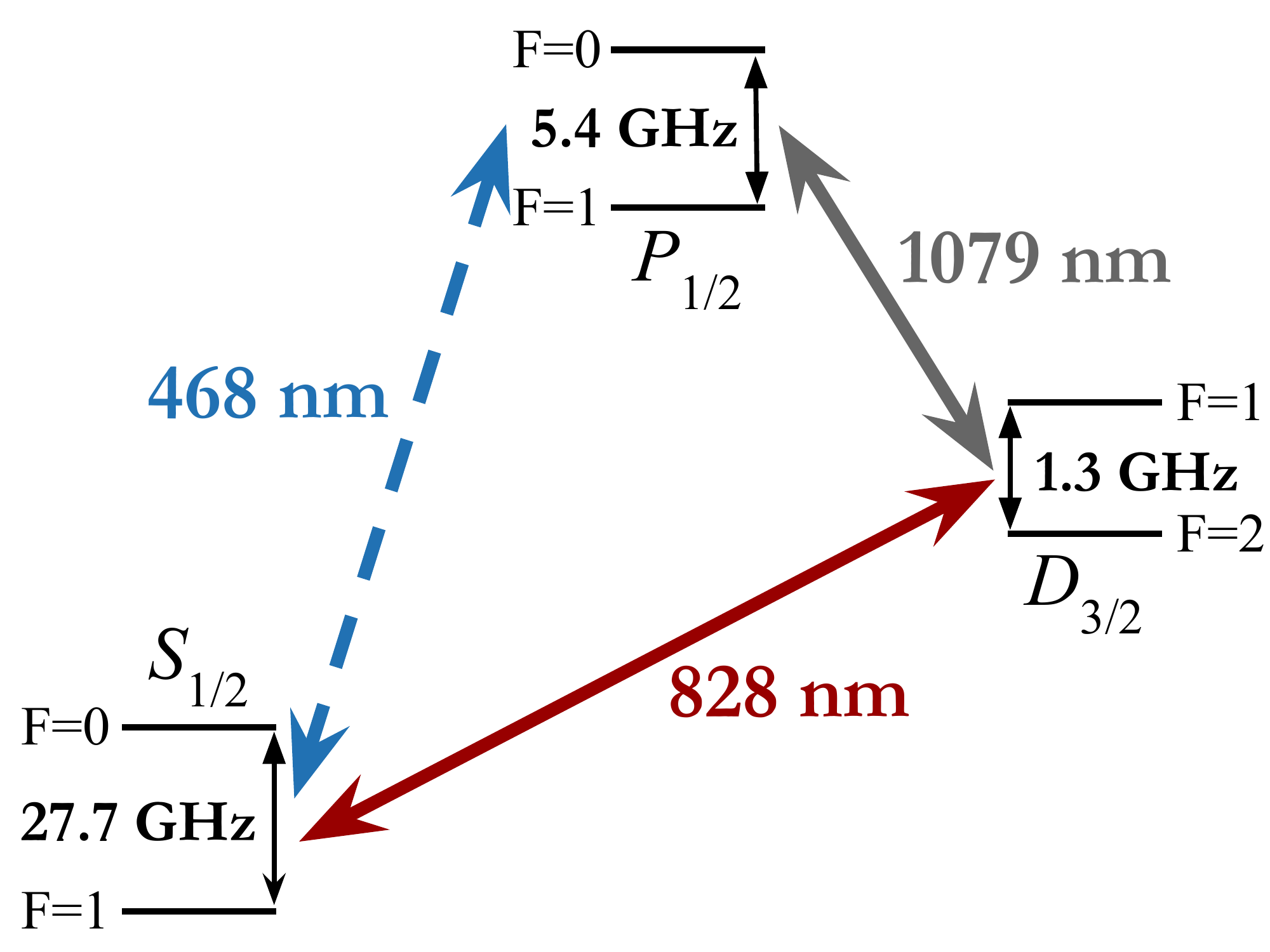}
    \caption{The $^{225}$Ra$^{+}$ level structure for clock operation showing the Doppler cooling (blue), repump (gray), and clock (red) transitions. Hyperfine splittings of the states are shown.  For two-laser clock operation the 468 nm laser is not required.}
    \label{fig:radium-225-hyperfine}
\end{figure}

Optical clocks using the $S_{1/2}\rightarrow D_{3/2}$ E2 transition in $^{171}$Yb$^+$ (nuclear spin $I=1/2$) have been implemented \cite{Stenger2001} and used to obtain the current best limit on the temporal drift of the fine structure constant \cite{Lange2021}. The hyperfine levels of the $^{171}$Yb$^+$ ion enable simple state preparation and detection by driving certain transitions between them \cite{Tamm2000}. The nonzero nuclear spin also provides first-order magnetic insensitive transitions at zero magnetic field, allowing for longer coherence times with no or less magnetic field stabilization. Among the alkaline earth elements, $^{225}$Ra$^+$ and $^{133}$Ba$^+$ also have a nuclear spin $I=1/2$ \cite{Fan2019, Hucul2017}, which allows a similar state preparation and detection method as $^{171}$Yb$^+$. In this work, we describe the operation of an alkaline earth ion $S_{1/2}\rightarrow D_{3/2}$ optical clock, and compare the performance and technical requirements of the $^{225}$Ra$^+$, $^{171}$Yb$^+$, and $^{133}$Ba$^+$ three-laser clocks.

\section{THREE-LASER ION CLOCK OPERATION}\label{sec:clock-operation}

\begin{figure}[b]
\centering{}
\includegraphics[height=5cm]{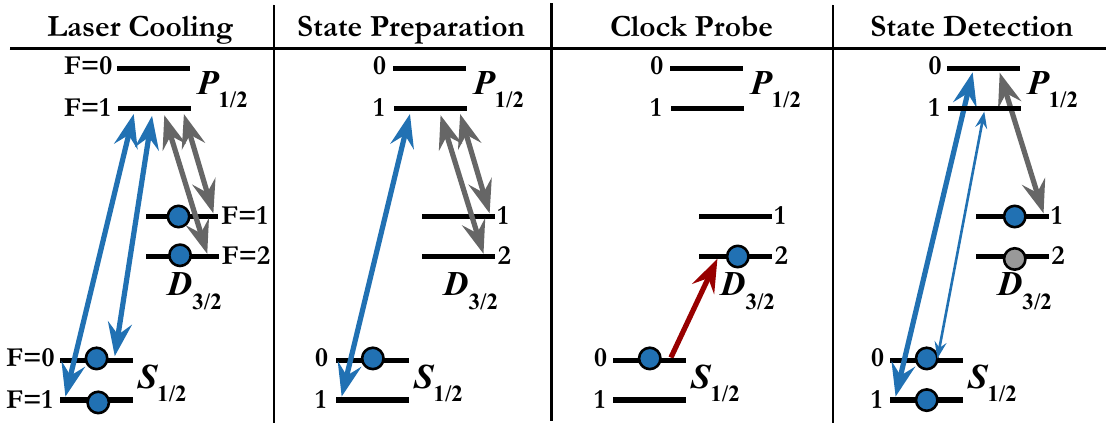}
    \caption{The clock measurement sequence for a three-laser $^{225}$Ra$^{+}$ ion optical clock is shown.  The clock laser drives the E2 $S_{1/2}\rightarrow D_{3/2}$ transition (red), Doppler cooling uses the $S_{1/2}\rightarrow P_{1/2}$ transition (blue), and repumping requires driving the $D_{3/2}\rightarrow P_{1/2}$ transition (gray).   Blue circles are populated hyperfine levels, and the gray circle indicates the dark state. The thin blue arrow in the State Detection panel is a transition driven with weak pumping light, as it only needs to be driven a few times before population in the $|S_{1/2}, F=0\rangle$ state is pumped out.}
    \label{fig:ra-clock-sequence}
\end{figure}

Four steps constitute a three-laser ion optical clock measurement sequence: laser cooling, state preparation, clock interrogation, and state detection, see Fig. \ref{fig:ra-clock-sequence}. We first discuss the sequence for alkaline earth ions, $^{225}$Ra$^+$ and $^{133}$Ba$^+$. During laser cooling, population is driven from all hyperfine levels in the $S_{1/2}$ and the $D_{3/2}$ states to the $|P_{1/2}, F=1\rangle$ level. Then turning off the light driving the $|S_{1/2},F=0\rangle\rightarrow |P_{1/2},F=1\rangle$ transition prepares the ion in the $|S_{1/2}, F=0, m=0\rangle$ sublevel of the ground state, where $F$ denotes the hyperfine level and $m$ denotes the magnetic sublevel. After state preparation, the $|S_{1/2}, F=0, m=0\rangle\rightarrow |D_{3/2}, F=2, m=0\rangle$ clock transition may be interrogated with either a Rabi or Ramsey sequence \cite{Letchumanan2004}. The $|S_{1/2},F=1\rangle\rightarrow |P_{1/2}, F=0\rangle$ and $|D_{3/2},F=1\rangle\rightarrow |P_{1/2}, F=0\rangle$ transitions are then used for state detection, while the $|S_{1/2},F=0\rangle\rightarrow |P_{1/2}, F=1\rangle$ transition is driven with weak pumping light.

If the ion is shelved in the $|D_{3/2},F=2,m=0\rangle$ sublevel by the clock interrogation pulse, the state detection light will be far off-resonant from transitions out of this dark state. If the ion remains in the $|S_{1/2},F=0,m=0\rangle$ sublevel, it is pumped into the cycling manifold by the weak pumping light, with a small probability of population leaking to the $|D_{3/2},F=2\rangle$ dark state. During state detection, scattered $S_{1/2}\rightarrow P_{1/2}$ photons are collected onto a detector.  If the total photon counts exceed a predetermined detection threshold, the state detection event is characterized as ``bright'', otherwise the event is ``dark''.  These bright and dark state detection events determine the probability that the clock transition was driven.

The operating procedure of the alkaline earth ion clock is similar to that of the Yb$^+$ E2 clock \cite{Tamm2000}, with the only difference being that the $^{171}$Yb$^+$ repump laser drives the $D_{3/2}\rightarrow D[3/2]_{1/2}$ transition (935 nm) instead of the $D_{3/2}\rightarrow P_{1/2}$ transition (2.4 $\mu$m) \cite{Bell1991}. For Yb$^+$, the $D_{3/2}\rightarrow D[3/2]_{1/2}$ transition is a more convenient wavelength and eliminates the coherent dark states formed by a superposition of the $S_{1/2}$ state and the $D_{3/2}$ state \cite{Berkeland2002}. A state with properties similar to Yb$^+$'s $D[3/2]_{1/2}$ state does not exist in alkaline earth ions.

\subsection{Quantum projection noise}

An important characteristic of an optical clock is the quantum projection noise (QPN)-limited instability. With a more stable clock, the same instability can be achieved in a shorter amount of averaging time. In general, the fractional instability of a clock is \cite{Peik2006}

\begin{equation}
    \Delta \nu(T) \propto \frac{1}{\mathrm{SNR} \sqrt{N T_\mathrm{p} T}},
\end{equation}
where SNR is the signal-to-noise ratio of the clock measurement in one cycle, $N$ is the number of atoms, $T_\mathrm{p}$ is the clock probe time, and $T$ is the total measurement time. For a typical single ion clock, $N=1$, and the QPN at a fixed measurement time is determined by the probe time and the SNR. The probe time affects the single-shot measurement linewidth, and is limited by the lifetime of the $D_{3/2}$ state. The state preparation and detection fidelity affects the signal to noise ratio of the clock signal, and is limited by off-resonant scattering and the branching fraction of the $P_{1/2}$ state to the $D_{3/2}$ state. The hyperfine splittings are $\sim 1$ GHz for the ions discussed in this work \cite{Hucul2017,MaartenssonPendrill1994, Neu1988}, and the corresponding off-resonant scattering rate is $\sim1\%$ for a experimental setup with a typical $0.1\%$ to $1\%$ photon collection efficiency \cite{Olmschenk2007}. The branching fraction of the $P_{1/2}$ to the $D_{3/2}$ state affects the SNR by optical pumping of the $|S_{1/2},F=0\rangle$ state to the $|D_{3/2},F=2\rangle$ state via the $|P_{1/2},F=1\rangle$ state, which has a probability of

\begin{equation}
    \epsilon = \frac{p_{\scaleto{DP}{4pt}}\times (1-p)}{1-p_{\scaleto{SP}{4pt}}\times p},
\end{equation}
where $p$ is the electronic branching fraction from the $P_{1/2}$ to the $S_{1/2}$ state, $p_{\scaleto{SP}{4pt}}=1/3$ is the hyperfine branching fraction from the  $|P_{1/2},F=1\rangle$ to the $|S_{1/2},F=0\rangle$ state, and $p_{\scaleto{DP}{4pt}}=5/6$ is the hyperfine branching fraction from the $|P_{1/2},F=1\rangle$ to the $|D_{3/2},F=2\rangle$ state. The lifetime and the electronic branching fraction of the ion species discussed in this work are listed in Table \ref{table:clock-elements}. For $^{133}$Ba$^+$ and $^{225}$Ra$^+$, the branching fraction induced error $\epsilon$ is much greater than the off-resonant scattering error, so the off-resonant scattering error can be ignored in these systems. Although the fidelities of the $^{225}$Ra$^+$ and the $^{133}$Ba$^+$ clocks are smaller than the $^{171}$Yb$^+$ clock, lower QPNs are achievable with longer probe times which are enabled by the longer $D_{3/2}$ state lifetimes. The branching fraction induced error also can be avoided if an additional microwave pulse is used for state detection, see Section \ref{subsec:two-laser-clock}.

\begin{table}
\centering{}
\caption{Relevant atomic properties for alkali-like ions are given: the $P_{1/2}$ state to $S_{1/2}$ state branching fraction $p$, transition wavelengths (nm), $D_{3/2}$ state lifetime $\tau_{3/2}$, and the $P_{1/2}$ state branching fraction induced error $\epsilon$. The repump transition for Yb$^+$ is the $D_{3/2}\rightarrow D[3/2]_{1/2}$ transition, while for Ba$^+$ and Ra$^+$ the $D_{3/2}\rightarrow P_{1/2}$ transition is used.}
\label{table:clock-elements}
\begin{tabular}{cccccccc}
    Ion & $p$ & $\lambda_\text{cool}$ & $\lambda_\text{repump}$ & $\lambda_\text{clock}$ & $\tau_{3/2}$ (s) & $\epsilon$ \\ 
    \hline \\ [-1.5ex]
    Ba$^{+}$ & 0.73182 \cite{Arnold2019b} & 493 & 650 & 2052 & 79.8(4.6) \cite{Yu1997} & 30\% \\
    Yb$^{+}$ & 0.995 \cite{Olmschenk2007} & 369.5 & $935^\ast$ & 435.5 & 0.052(1) \cite{Gerz1988} & $<1\%$\\
    Ra$^{+}$ & 0.9104 \cite{Fan2019} & 468 & 1079 & 828 & 0.638(10) \cite{Pal2009} &  11\% \\
\end{tabular}
\end{table}

\section{Two Laser Clock Operation}\label{subsec:two-laser-clock}

\begin{figure}
\centering{}
\includegraphics[height=5cm]{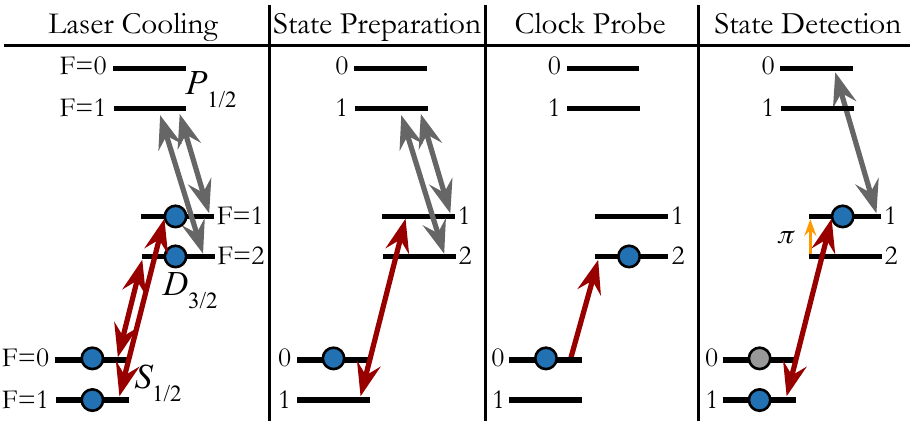}
    \caption{Clock measurement sequences for a two-laser $^{225}$Ra$^{+}$ ion optical clock based on the $S_{1/2}\rightarrow D_{3/2}$ transition (red), which is also used for Doppler cooling.  Arrows represent transitions between hyperfine levels, blue circles represent populated hyperfine levels, and gray circles represent the dark state during state detection.  The $D_{3/2}\rightarrow P_{1/2}$ repump transition (gray) and $|D_{3/2}, F=2, m=0\rangle\rightarrow |D_{3/2}, F=1, m=0\rangle$ microwave transition (yellow) are shown.}
    \label{fig:microwave-sequence}
\end{figure}

A three-level clock may also enable clock operation with only two lasers addressing two electronic transitions.  The ion could be laser cooled by strongly driving the $S_{1/2}\rightarrow D_{3/2}$ electric quadrupole transition and repumping with the $D_{3/2}\rightarrow P_{1/2}$ transition.  Laser cooling has been achieved by driving the $S_{1/2}\rightarrow D_{5/2}$ electric quadrupole transition in $^{40}$Ca$^{+}$ \cite{Hendricks2008}.   For a three-level clock cooling on the $S_{1/2}\rightarrow D_{3/2}$ transition drops the requirement to drive the $S_{1/2}\rightarrow P_{1/2}$ transition.  For $^{225}$Ra$^+$ the clock would only need 828 nm and 1079 nm lasers, see Fig. \ref{fig:microwave-sequence}.  To perform Doppler cooling the $|S_{1/2}, F=0\rangle\rightarrow | D_{3/2}, F=2\rangle$ and $|S_{1/2}, F=1\rangle\rightarrow |D_{3/2}, F=1\rangle$ transitions are driven while driving the $|D_{3/2}, F=2\rangle\rightarrow |P_{1/2}, F=1\rangle$ and $|D_{3/2}, F=1\rangle\rightarrow |P_{1/2}, F=1\rangle$ transitions to repump population.  The population is then prepared in the $|S_{1/2}, F=0, m=0\rangle$ state by turning off the light driving the $|S_{1/2}, F=0\rangle\rightarrow | D_{3/2}, F=2\rangle$ transition.  For clock interrogation, the $|S_{1/2}, F=0, m=0\rangle\rightarrow |D_{3/2}, F=2, m=0\rangle$ transition is driven, followed by a $|D_{3/2}, F=2, m=0\rangle\rightarrow |D_{3/2}, F=1, m=0\rangle$ microwave $\pi$-pulse at 1.3~GHz which populates a bright state if the clock transition was driven.  The $|S_{1/2}, F=1\rangle\rightarrow |D_{3/2}, F=1\rangle$ and $|D_{3/2}, F=1\rangle\rightarrow |P_{1/2}, F=0\rangle$ transitions are then driven for state detection.  Higher intensity would be needed at 828 nm for laser cooling compared to the three laser scheme described in Sec. \ref{sec:clock-operation}, but the trade-off in power could be enabling for integrated photonics as the only required wavelengths are now in the infrared.

\section{Conclusion}

We have proposed three-level clocks with alkali-like ions with non-zero nuclear spin based on the $S_{1/2}\rightarrow D_{3/2}$ transition. We show that species with nuclear spin $I=1/2$, including $^{133}$Ba$^{+}$, $^{171}$Yb$^{+}$, and $^{225}$Ra$^{+}$, are promising for realizing such clocks.  We highlight $^{225}$Ra$^{+}$ as a candidate system for a three-level transportable clock due to its low state detection infidelity, ability to achieve low total systematic uncertainty, and suitability for integrated photonics \cite{Holliman2022}.  Optical losses in integrated photonics generally decrease the farther the light is from the UV \cite{Liu2018, West2019}, which makes clock species such as Ra$^+$ \cite{Holliman2022}, whose shortest wavelength is at 468~nm, promising for use with integrated photonics.  The radium ion clock $\lambda=828$~nm and repump $\lambda=1079$~nm transitions could also be frequency summed to generate the $\lambda=468$~nm Doppler cooling light, eliminating the need for a blue laser.

\begin{table}[b]
\centering{}
\caption{Candidate alkali-like isotopes and their relevant atomic properties are given, including the half-life, the evaluated $|S_{1/2}, F=0, m=0\rangle\rightarrow |D_{3/2}, F=2, m=0\rangle$ clock transition and quadratic Zeeman (QZ) shift at zero magnetic field, the differential scalar polarizability $\alpha_0$ (atomic units) of the $S_{1/2}\rightarrow D_{3/2}$ clock transition, and the magic rf trap drive frequency $\Omega_\text{rf}$.}
\label{table:clock-species}
\begin{tabular}{cccccccc}
    Species & Half-life & QZ (Hz/G$^2$) & $\alpha_{0}$ & $\Omega_\text{rf}$ (MHz) \\ 
    \hline \\ [-1.5ex]
    $^{133}$Ba$^{+}$ & 10.5~y &-532 & -75.4(1.1) \cite{Sahoo2009a}   & 3.4\\
    $^{171}$Yb$^{+}$ & stable & 519  & 42(8) \cite{Schneider2005}  & -\\
    $^{225}$Ra$^{+}$ & 14.9~d &-321 & -20.8(1.7) \cite{Sahoo2009a}  & 6.0\\
\end{tabular}
\end{table}

Among the alkaline-earth elements, the radium ion has the lowest sensitivity to blackbody radiation, and is promising for reaching total systematic uncertainty at the low $10^{-18}$ level \cite{Versolato2011b, Holliman2022}. Similar to the other alkali-like ions, the radium ion's $S_{1/2}\rightarrow D_{3/2}$ clock transition has a negative differential scalar polarizability.  This enables operation at a magic rf trap drive frequency, see Table \ref{table:clock-species}, such that the micromotion induced-scalar Stark shift and the second-order Doppler shift cancel \cite{Dube2014}. For $^{225}$Ra$^{+}$, the quadratic Zeeman shift coefficient at zero field is smaller than that of the $^{171}$Yb$^{+}$ ion, see Table \ref{table:clock-species}.  In a clock frequency comparison using the proposed protocol Ra$^{+}$ could help improve the constraints on $\dot\alpha/\alpha$ \cite{Lange2021}.  For Ra$^{+}$, the sensitivity to the time variation of the fine structure constant $\dot\alpha/\alpha$ of the $S_{1/2}\rightarrow D_{3/2}$ clock transition, $K_{3/2}=3.03$, is higher than the $S_{1/2}\rightarrow D_{5/2}$ transition, $K_{5/2}=2.78$, which is the largest positive sensitivity among demonstrated optical clocks \cite{Flambaum2009}. This increase in sensitivity also comes with an improved stability because of the longer $D_{3/2}$ state lifetime, 638(10)~ms, compared to the $D_{5/2}$ state lifetime, 303(4)~ms \cite{Pal2009}.

\acknowledgments       
 
We thank Samuel Brewer and Shimon Kolkowitz for feedback on the manuscript, and David Leibrandt for helpful discussions and pointing out reference \cite{Hendricks2008}. This research was performed under the sponsorship of ONR N00014-21-1-2597, DoE DE-SC0022034, and NSF PHY-2146555.

%\bibliography{references}
\bibliographystyle{spiebib}

\end{document}